\documentclass{osa-article}
\usepackage{graphicx}
\usepackage{dcolumn}
\usepackage{bm}
\usepackage{epstopdf}

\journal{oe}

\articletype{Research Article}

\begin{document}

\title{Characterization of Suspended Membrane Waveguides towards a Photonic Atom Trap Integrated Platform}

\author{
Michael Gehl,\authormark{1,$\dagger$}
William Kindel,\authormark{1,$\dagger$}
Nicholas Karl,\authormark{1}
Adrian Orozco,\authormark{1,2}
Katherine Musick,\authormark{1}
Douglas Trotter,\authormark{1}
Christina Dallo,\authormark{1}
Andrew Starbuck,\authormark{1}
Andrew Leenheer,\authormark{1}
Christopher DeRose,\authormark{1}
Grant Biedermann,\authormark{3}
Yuan-Yu Jau,\authormark{1} and
Jongmin Lee\authormark{1,*}}

\address{
\authormark{1}Sandia National Laboratories, Albuquerque, New Mexico 87185, USA\\
\authormark{2}Department of Physics and Astronomy, University of New Mexico,  Albuquerque, NM 87106, USA\\
\authormark{3}Department of Physics and Astronomy, University of Oklahoma, Norman, Oklahoma 73019, USA}

\email{\authormark{*}jongmin.lee@sandia.gov} 



\begin{abstract}
We demonstrate an optical waveguide device, capable of supporting the high, in-vacuum, optical power necessary for trapping a single atom or a cold atom ensemble with evanescent fields. Our photonic integrated platform, with suspended membrane waveguides, successfully manages optical powers of 6 mW (500\,$\mu$m span) to nearly 30 mW (125\,$\mu$m span) over an un-tethered waveguide span.  This platform is compatible with laser cooling and magneto-optical traps (MOTs) in the vicinity of the suspended waveguide, called the membrane MOT and the needle MOT, a key ingredient for efficient trap loading $-$ \cite{Lee20, Jau17}.  We evaluate two novel designs that explore critical thermal management features that enable this large power handling. This work represents a significant step toward an integrated platform for coupling neutral atom quantum systems to photonic and electronic integrated circuits on silicon.
\end{abstract}

\section{Introduction}

\begin{figure}[b!]
\centering\includegraphics[width=0.6\columnwidth]{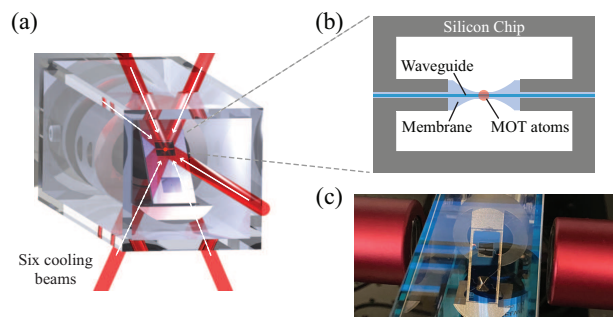}
\caption{The concept of a photonic atom trap integrated platform (ATIP). (a) 3D render of six-beam MOT on the integrated photonic platform in the vacuum chamber. (b) Conceptual drawing of an integrated photonic platform for atom-light interaction. (c) Picture of an integrated photonic platform in vacuum. Light is coupled in and out through high numerical aperture (NA) lenses.}
\label{fig_1}
\end{figure}

Over the past 20 years there have been remarkable advances in quantum computing and quantum sensing using neutral atoms\,\cite{Weiss17, Alzar19}. With the advent of microfabrication techniques, new frontiers in quantum applications\,\cite{DiVincenzo00, Zoller01, Qi18} have been explored  through atom-light interaction\,\cite{Mabuchi06, Stern13, Rolston13a, Lukin14, Kimble15, Rolston15, Stievater16, Black18, Pfau18, Hung19, Hackermueller20, Kimble20, Ayi-Yovo20} in addition to incorporating compact MOTs\,\cite{Takuma91, Jhe96, Hinds09, Arnold10, Hinds13, Lee13, Bongs16, Squires17, Arnold17} on atom chips\,\cite{Hansch99, Schmiedmayer00, Folman05, Anderson05, Reichel07, Anderson13,Himsworth14, Folman16} and superconducting circuits\,\cite{Schmiedmayer09, Rolston11, Rolston13b, Schmiedmayer15, Park15, Fortagh17}. For quantum applications, integrated photonic circuits using silicon photonics, III-V photonics and nonlinear optics are engineered to demonstrate compact on-chip laser systems\,\cite{Kodigala19, Zwiller20, Hazan20}. Such photonic technologies also expand reliable operation of cold atom positioning, navigation and timing (PNT) sensors\,\cite{Biedermann14} in a dynamic environment and enable an integrated on-chip quantum computing platform capable of individual spin addressing, spin-spin entanglement and spin readout. In addition, quantum processers leveraged by superconducting circuits and integrated electronics have been demonstrated\,\cite{Knight17, Martinis19}, and novel surface ion-trap platforms\,\cite{Zhang17, Chiaverini20} have been explored through the on-chip integration of microfabricated surface electrodes and integrated photonics. All these quantum engineering efforts converge to realize a true quantum device by utilizing classical computing and control systems through a quantum-to-classical interface\,\cite{Reilly15}.

Integrated photonic waveguide platforms have been explored for over a decade to realize trapped neutral atoms via evanescent light. This waveguide platform takes advantage of the small mode area of the waveguide which enables strong atom-light interaction, enhanced photon collection from atoms, low SWaP (size, weight and power) operation, as well as chip-scale design and flexibility. Unlike the low absorption loss of optical fibers, the higher absorption losses of integrated photonic waveguides make it difficult to sufficiently dissipate the heat generated by the laser power needed to trap the neutral atoms in a vacuum. Here, we demonstrated sufficient in-vacuum power handling capability at the integrated suspended waveguide, which will enable sufficient trapping power and lead to a first demonstration of trapped neutral atoms. This unique photonic platform is based on a membrane ridge waveguide\,\cite{Ayi-Yovo20} which is fully suspended across the entire silicon chip and has lengths of several centimeters with no defects and high optical transmission. In order to create MOT atoms that surround a portion of the waveguide, an opening is required through the silicon substrate that allows the free space cooling beams to propagate.

The integrated photonic waveguide platform with trapped neutral atoms can be an exemplary technique for integrated on-chip solutions for quantum applications. The integrated photonic platform can manipulate the trapped neutral atoms in a manner consistent with scalable, integrated quantum systems. The fundamental power of integrated quantum systems is routinely demonstrated with artificial solid-state spins in cryostats. However, among spin candidates, atomic spins of neutral atoms have superior homogeneous physical properties and long coherence times\,\cite{Kaufman20} compared to artificial counterparts\,\cite{Englund16,Popkin16}. This led to the recent successful demonstration of high-fidelity two-qubit gate operation with Rydberg electric dipole-dipole interactions\,\cite{Jau16, Lee17, Omran19, Leseleuc19, Grahm19, Endres20}. 

In this paper, we demonstrate the integrated photonic platform with alumina membrane and silicon needle structures toward the photonic atom trap integrated platform (ATIP), which is designed to perform atom-light interaction with the membrane ridge waveguide\,\cite{Ayi-Yovo20} as shown in Fig.\,\ref{fig_1}. Analogous to a nanofiber evanescent field optical trap\,\cite{RB10, Lee15, Polzik17, Aoki19}, the integrated photonic platform can trap, control, and probe neutral atoms through the evanescent field of optical waveguide modes, and the heat generated from optical absorption in the waveguide can be dissipated through the membrane and needle structures attached to the substrate. The primary obstacles for photonic ATIP demonstration are (1) efficient atom loading into the trap, located within hundreds of nanometers proximity to the micro- and nano-structures and (2) heat dissipation of the suspended waveguide in vacuum. These issues define a barrier to entry for realizing an integrated evanescent-field optical trap (EFOT) and have been studied for more than a decade in the field\,\cite{Lee20, Jau17, Rolston15, Stievater16, Black18, Kimble20}. With this work, we demonstrate an approach that crosses this threshold and can be an enabling technology for a new breed of quantum devices. 

\section{Optical Waveguides on Membrane and Needle Structures}

\begin{figure}[b!]
\centering\includegraphics[width=0.6\columnwidth]{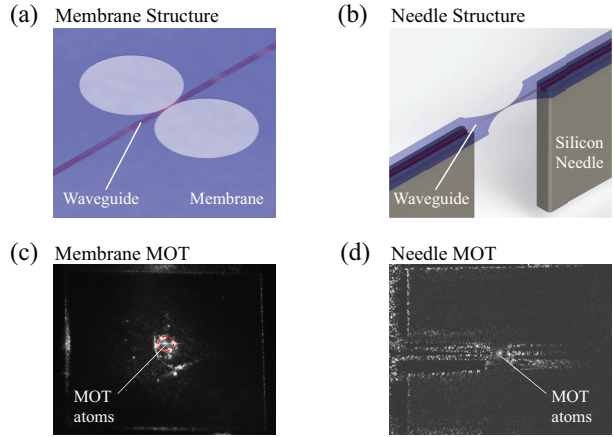}
\caption{(a) 3D render of ``membrane structure''. The waveguide (red line) passes the open hole that is needed for atom loading. Membranes efficiently dissipate heat from the suspended waveguide to the silicon substrate in vacuum. (b) 3D render of ``needle structure''.  The waveguide (red line) passes across an opening in the substrate that is beneficial for atom loading. Silicon needles connected to the silicon substrate efficiently dissipate heat from the suspended waveguide in vacuum. (c) Image of ``membrane MOT'' atoms on a straight dummy waveguide with the width of 3\,$\mu$m across a membrane hole\,\cite{Lee20}, where the membrane hole is specified with the red-dashed circle and the waveguide span is 400\,$\mu$m  (see Fig.\,\ref{fig_6}(b) straight design). The material of the membrane and waveguide is silicon nitride. (d) Background subtraction image of ``needle MOT'' atoms on the membrane ridge waveguide. The span is 400\,$\mu$m (see Fig.\,\ref{fig_6}(c) hybrid needle design); the gap between two silicon needles is 610\,$\mu$m. The material of the membrane and waveguide is alumina.}
\label{fig_2}
\end{figure}

To develop the integrated photonic waveguide platform, we used successful features and learned about the constraints from earlier designs\,\cite{Lee20, Jau17} such as ``membrane structure'' and ``needle structure'' which show that sub-Doppler cooling with these geometries works well with or without a dummy waveguide. A Sub-Doppler cooling process\,\cite{Dalibard89, Chu89, Metcalf90, Phillips92, Foot92, Hinds11} will be required to load atoms into the EFOT. We can create a MOT around the suspended waveguide in sub-millimeter openings of ``membrane structure'' and ``needle structure'' as shown in Fig.\,\ref{fig_2}. The MOT using these structures is called ``membrane MOT'' and ``needle MOT''. In the first ``membrane structure'', the transparent membrane creates two large, separate hemispherical capture volumes to load cold atoms into the sub-millimeter membrane hole. In the second ``needle structure'', analogous to a six-beam MOT, an etched opening in a silicon substrate creates a large, unimpeded, capture volume to load cold atoms into a sub-millimeter gap between two silicon needles. Importantly, both structures enable concentrating atoms near the fabricated nano-structures while simultaneously overcoming the limited atomic-position-accessibility resulting from atomic collisions at the interface of micro- and nano-structures. Both the membrane and needle structures are integrated with optical waveguides, and also serve to dissipate the heat generated from optical absorption in the suspended waveguide. When under vacuum, the heat dissipation across a sub-millimeter atom loading zone, such as a membrane hole or a gap between two silicon needles, is limited to radiation or conduction through the membrane or needle structures. This capacity for heat dissipation limits the maximum optical power handling capability that the waveguide can support. In particular, longer suspended waveguides enable efficient atom loading, while shorter suspended waveguides increase heat dissipation capacity. In this paper, we demonstrate a needle MOT around an alumina membrane suspended waveguide with a hybrid needle device as shown in Fig.\,\ref{fig_2}(d), the design of which is similar to the optical devices used for the heat dissipation test.

By using the capability of efficient atom loading at a sub-millimeter opening, this structure will enable loading of atoms into the EFOT after sub-Doppler cooling, which can be coupled to monolithically integrated photonic and electronic integrated circuits, a major step towards a standalone photonic ATIP. Such a system can utilize silicon photonics (low-loss photonic waveguides, modulators/switches, optical filters, and Ge photodiodes) and III-V compound semiconductor photonics (tunable DBR lasers, SOA, modulators, and photodiodes) with CMOS RF/analog/digital integrated circuits.

\section{Design of Integrated Photonic Waveguide Platforms for Atom-Light Interaction}

\begin{figure}[b!]
\centering\includegraphics[width=0.6\columnwidth]{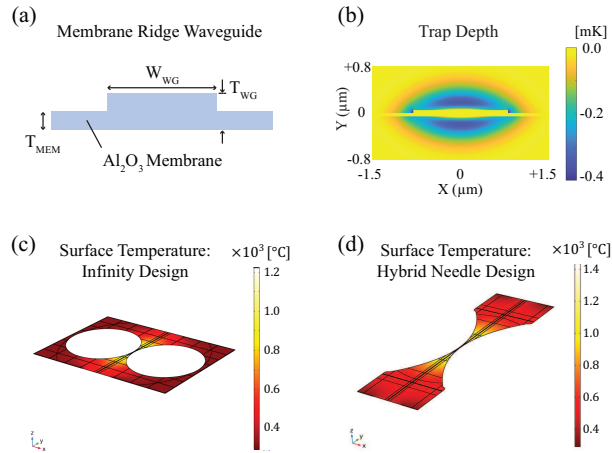}
\caption{Design of a photonic ATIP. (a) Alumina ($\rm{Al_{2}O_{3}}$) ridge membrane waveguide. $\rm{W_{WG}}$ is the width of the waveguide, $\rm{T_{WG}}$ is the thickness of the ridge waveguide structure, and $\rm{T_{MEM}}$ is the thickness of the membrane. This is designed to guide three wavelengths of 793\,nm, 852\,nm, and 937\,nm for trapping and probing Cs atoms. (b) Trapping potential calculation ($\sim$350\,$\mu$K) with 793\,nm (3.27\,mW) and 937\,nm (2.73\,mW) light, i.e.,  two magic-wavelength blue- and red-detuned traps for Cs atoms. This EFOT is based on two quasi-TE modes of the membrane waveguide ($\rm{W_{WG}}$ = 1.6\,$\mu$m, $\rm{T_{WG}}$ = 100\,nm, and $\rm{T_{MEM}}$ = 50\,nm). (c) Heat dissipation simulation of the ``infinity design'' structure. (d) Heat dissipation simulation of the ``hybrid needle design'' structure.}
\label{fig_3}
\end{figure}

For the design of a membrane ATIP, we need to consider several design parameters. First of all, we choose alumina, i.e., aluminum oxide ($\rm{Al_{2}O_{3}}$), as the material of the membrane and waveguide. Alumina has been demonstrated as an excellent, low loss,  waveguide material for wavelengths from UV to NIR\,\cite{Aslan10, West19, Agaskar19}.  Furthermore, alumina has the advantage of increased resistance to alkali vapor. By fabricating the waveguide entirely from alumina, we can avoid the additional process of alumina coating on silicon nitride (SiN) or aluminum nitride (AlN) waveguides proposed by some published papers\,\cite{Pfau18}. Further design parameters, such as the thickness of the membrane and membrane waveguides (see Fig.\,\ref{fig_3}(a)), will affect (1) the transmission of cooling beams through the membrane, (2) the trap depth of the EFOT, and (3) the heat dissipation through the membrane. Obtaining maximum transmission through the alumina membrane ($\rm{n_{\rm{Al_{2}O_{3}}}=1.76}$) is critical for efficient atom loading and cold dense atoms around the waveguides. For a membrane of alumina, surrounded by vacuum, a thickness near 265\,nm produces an anti-reflection minima, leading to $> 98\%$ transmission for 852\,nm cooling beams of Cs atoms, however this thickness is not optimal for the EFOT. To maximize the evanescent field extending from the optical waveguide, we explore waveguide thicknesses of 75, 100, 125 and 150\,nm, and the membrane thickness is chosen to be thinner than the waveguide to form a ridge waveguide structure (Fig.\,\ref{fig_3}(a)).  We have investigated membrane thickness of 25, 50 and 75\,nm, leading to, respectively, $96.2\%$, $88.0\%$ and $80.3\%$ transmission for the circularly polarized cooling beams.  For simplicity of fabrication, the entire membrane is a constant thickness.  In future iterations, additional deposition or etching steps can be added in order to produce a thicker, anti-reflection membrane, while still maintaining a thin ridge waveguide.

The range of waveguide and membrane thicknesses explored was based on optimization of the evanescent trap potential.  The thicknesses, along with the waveguide width is optimized to provide the maximum trap potential per mW of optical power within the waveguide.  Fig.\,\ref{fig_3}(b) shows an example of the calculated trap potential for a certain waveguide geometry, demonstrating 350\,$\mu$K trap depth with 6\,mW of total optical power. A further benefit of this thin waveguide geometry is that the optical mode is weakly confined to the dielectric material. This provides reduced optical absorption, reducing the heat generated within the waveguide.

In this paper we explore two primary designs. The first, named the ``infinity design'', is illustrated in Fig.\,\ref{fig_2}(a), and the second design, named ``hybrid needle design'', is shown in Fig.\,\ref{fig_2}(b). Both designs are based on alumina membranes on a silicon substrate. The ``infinity design'' features a continuous membrane which extends across the large cooling beam window which is opened in the silicon substrate.  A ridge waveguide can be implemented on the thin membrane film, passing across the window, from one side of the chip to the opposite side. At the center of the cooling beam window, two circular openings are formed in the membrane, on opposite sides of the waveguide. This provides large open area for formation of the membrane MOT, while providing sufficient thermal dissipation of heat generated in the waveguide at the contact of the two circular openings. For the ``hybrid needle design'', the etched cooling beam window retains two small silicon needles which extend across the center from opposite sides (see the device on the right side in Fig.\,\ref{fig_4}(b)). These needles support the membrane near the membrane waveguide, providing improved thermal dissipation. In the end, most of the membrane is removed from the cooling beam window.  This is expected to improve the atom capture into the needle MOT, however, the silicon needles can partially block and reduce cooling beam transmission.  Both devices feature a 6\,mm by 6\,mm window through the silicon substrate.  In Fig.\,\ref{fig_3}(c-d), we show COMSOL simulations of the temperature distribution of both designs in the vicinity of the atom loading region when 10\,mW of optical power is passing through the waveguide. We simulate the device using a conservative value of thermal conductivity of $1\,\rm{[W/m \cdot K]}$ for atomic layer deposited alumina thin films on silicon\,\cite{Hopkins18}.  By coupling light into the waveguide and imaging from above, we can observe the intensity of scattered light along the length of the waveguide.  By fitting the decay of this intensity to an exponential, we estimate the waveguide to have a total propagation loss of $1.0 \pm 0.1\,\rm{dB/cm}$, due to combined absorption and scattering. We find excellent agreement between our thermal simulations and experiment if the optical absorption of the waveguide is chosen to be $1.0 \,\rm{dB/cm}$, suggesting that the measured propagation loss is largely due to material absorption, rather than scattering.  We find in simulation a peak temperature of approximately 1200 K for the ``infinity design'' and 1400 K for the ``hybrid needle design'', well below the melting point of alumina of 2354 K.  In addition to the full sized devices pictured in Fig.\,\ref{fig_4}(b), we fabricated an array of devices with varying suspended waveguide lengths in order to study the variation in optical power handling and to calibrate our thermal modeling. The results of measurements on these devices are discussed in section 5.

\section{Fabrication}

\begin{figure}[b!]
\centering\includegraphics[width=1\columnwidth]{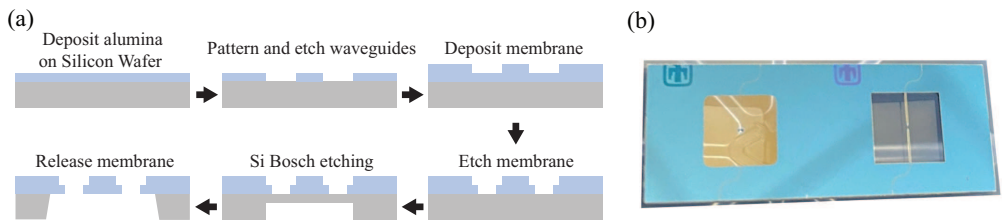}
\caption{(a) Fabrication process of a photonic ATIP.  Beginning with bulk silicon wafers, an alumina film is deposited by atomic layer deposition.  The film is patterned into waveguides, followed by deposition of a second film which acts as the membrane.  The membrane is patterned and etched.  A deep silicon Bosch etch is used to open cooling beam windows in the substrate, followed by a finishing $\rm{XeF_2}$ etch to suspend the waveguide regions. (b) Image of a fully fabricated ATIP, featuring the ``infinity design'' on the left and the ``hybrid needle design'' on the right.  Both designs feature a 6\,mm x 6\,mm cooling beam window through the silicon substrate.}
\label{fig_4}
\end{figure}

The devices described here are fabricated in a novel suspended membrane waveguide platform based on low optical loss, alumina dielectric deposited by atomic layer deposition.  Unlike most integrated photonic waveguide platforms which are deposited on a silicon oxide cladding, our waveguides are deposited directly on a silicon substrate.  Silicon provides two orders of magnitude improvement in thermal conduction over silicon dioxide.  During fabrication, openings are etched in the alumina membrane along each side of the optical waveguide.  This allows the silicon to be etched away locally, creating a trench in the silicon which follows underneath the waveguide.  The implementation of a trench is critical to prevent optical absorption into the silicon substrate.  This unique platform creates a waveguide which is fully suspended across the entire chip.  We have demonstrated waveguides with lengths of several centimeters fully suspended, with no defects and high optical transmission. In order to create a MOT which surrounds a portion of the waveguide, we require an opening through the silicon substrate which allows the free space cooling beams to propagate.  This is achieved by a deep backside etch into the silicon substrate.  We typically open windows greater than 5\,mm on each edge. 

The fabrication process of the membrane ATIP is shown in Fig.\,\ref{fig_4}. Tensile alumina films are deposited on the front side of a silicon wafer via atomic layer deposition (ALD). The front side is patterned by conventional photolithography with photoresist (no hard mask). Alumina films are patterned and etched by inductively coupled plasma reactive-ion etching (ICP-RIE) to define the waveguide and membrane geometries. To form cooling beam windows, the back-side Si wafer is etched using a deep reactive ion etch (DRIE) Bosch process, stopping short of the alumina membrane by a few tens of microns. A final $\rm{XeF_2}$ etch is used to completely open these windows.  At the same time, in regions outside of the large cooling beam windows, small membrane openings along the length of the waveguide allow local undercutting of the waveguide.  This creates a trench in the silicon, on the order of 50um in width, completely suspending the alumina waveguide. In the  fabrication process, the thickness of the membrane and the dimension of waveguides is chosen (see Fig.\,\ref{fig_3}(a)) to obtain low propagation loss and create sufficient EFOTs (see Fig.\,\ref{fig_3}(b)); we have demonstrated devices with membrane thickness, $\rm{T_{MEM}}$, of 25, 50 and 75\,nm, and waveguide thickness, $\rm{H_{WG}}$, of 75, 100, 125 and 150\,nm.  
 
\section{Experimental Setup and Measurement}

\begin{figure}[b!]
\centering\includegraphics[width=1\columnwidth]{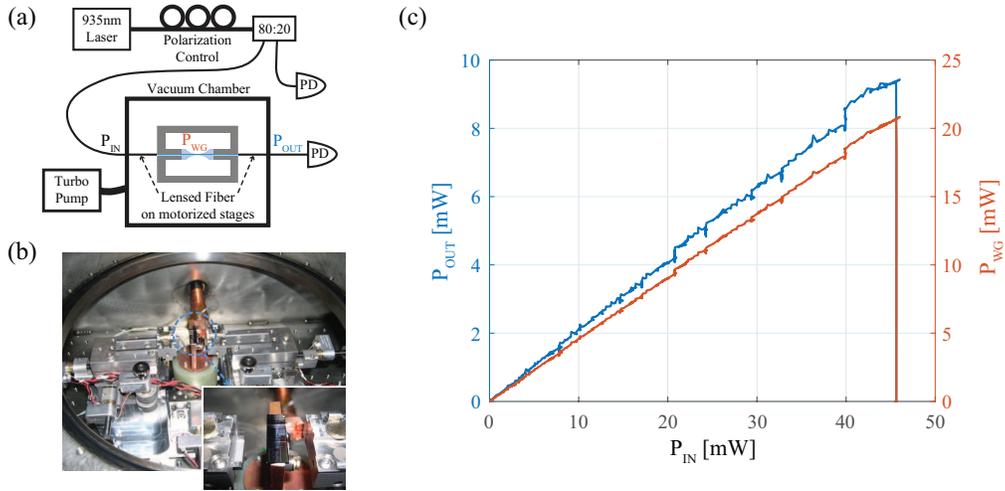}
\caption{Heat dissipation test results of a membrane ATIP in a $5.5 \times 10^{-5} \,\rm{torr}$ vacuum. (a) Schematic of the experimental setup, where ``80:20'' is a 80:20 fiber beam splitter; ``PD'', a photodiode; $P_{IN}$, the input power to the device; $\rm{P_{WG} = \sqrt{P_{IN} \cdot P_{OUT}}}$, the waveguide coupled power; $\rm{P_{OUT}}$, the output power from the device. (b) Image of a device sample on the vacuum-compatible piezoelectric motorized stages. (c) $\rm{P_{IN}}$ vs $\rm{P_{OUT}}$ and $\rm{P_{IN}}$ vs $\rm{P_{WG}}$ for the ``hybrid needle design'' with the span of 250\,$\mu$m, where the span corresponds to the length of the tapered membrane between two silicon needles as shown in Fig.\,\ref{fig_2}(b); the gap between two silicon needles is 460\,$\mu$m.}
\label{fig_5}
\end{figure}

A crucial metric of this design is the heat dissipation it provides. To characterize this quantity, we perform experimental measurements in which we test devices to failure. As shown in Fig.\,\ref{fig_5}(a-b), we used a custom vacuum chamber comprised of vacuum-compatible piezoelectric motorized stages for precise fiber-to-waveguide light coupling, and a vacuum-compatible temperature-controlled sample stage. The top of the chamber has an optical window sealed with an O-ring that limits the vacuum level. This vacuum chamber is connected to a turbo pump, and we can maintain a vacuum of $10^{-6}$ to $10^{-5}$ torr, which is sufficient to test the heat dissipation of the device samples in a convection-less environment (no air). First, we install the device sample on the sample stage, and mount two 780HP lensed-fibers (or cleaved fibers) on the fiber mount stages. By using the piezoelectric motorized stages, we first verify light coupling into the sample in ambient conditions. After closing the O-ringed optical window, the vacuum chamber becomes evacuated by the turbo pump, and we test the device under vacuum. We can simultaneously test multiple devices, without breaking vacuum, by using the vacuum compatible piezoelectric motorized stages.

As shown in Fig.\,\ref{fig_5}(a), we used a laser system comprised of a 935\,nm external cavity diode laser (ECDL), an optical isolator, and a tapered amplifier, which provides several hundred milli-watts of optical power. The fiber-coupled light passes through polarization control fiber paddles because the membrane waveguides prefers the quasi-TE mode. After this, the input power, $\rm{P_{IN}}$, was monitored with a 80:20 fiber splitter and a photodiode. Light which transmitted through the waveguide under test was collected in fiber and monitored by a second photodiode ($\rm{P_{OUT}}$). By assuming equal fiber to waveguide coupling on both sides of the device, we can estimate the waveguide-coupled power as $\rm{P_{WG} = \sqrt{P_{IN} \cdot P_{OUT}}}$, based on the input power and the output power as shown in Fig.\,\ref{fig_5}(c). Over testing many devices, we observe consistent transmission of $\sim20\%$, which gives confidence in this assumption.

\begin{figure}[b!]
\centering\includegraphics[width=1\columnwidth]{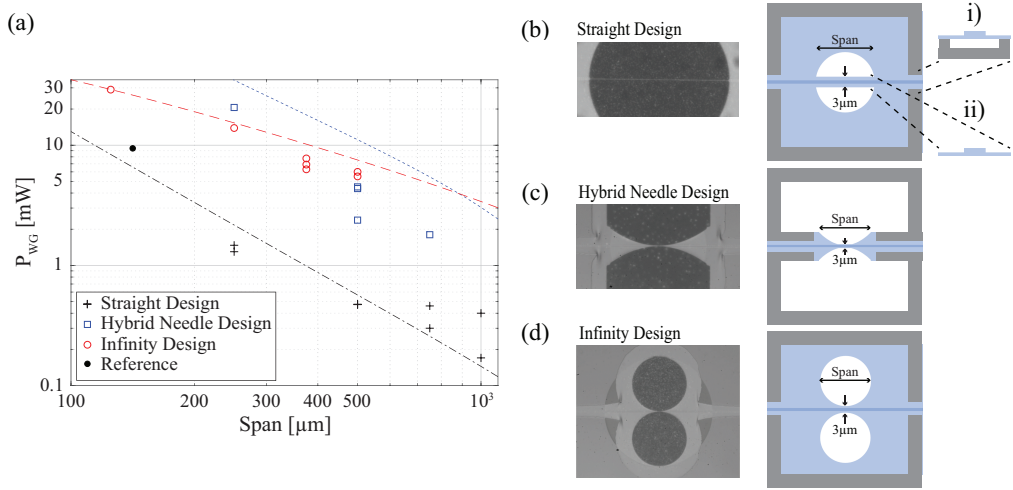}
\caption{(a) Measured maximum optical powers in the suspended membrane waveguides at failure and heat dissipation simulations for ``straight design'' (black cross and black-dash-dot line),  ``hybrid needle design'' (blue square and blue-dot line), and ``infinity design'' (red circle and red-dash line), respectively. The optical absorption of alumina waveguides is chosen to be $1.0 \,\rm{dB/cm}$ for finding excellent agreement between our thermal simulations and experiment. The black circle represents the result in a reference\,\cite{Kimble20}, which dissipates heat from silicon nitride waveguides using periodically spaced thermal anchoring structures. (b)  ``Straight design'' is based on the membrane structure. (Left) Picture of a fabricated device. (Right) Conceptual drawing of ``straight design''. i) (Cross-section view) The membrane waveguide above a trench undercut. ii) (Cross-section view) The suspended membrane waveguide. The span is the length of the suspended membrane waveguide.  (c) ``Hybrid needle design'' is based on the membrane and needle structures. (Left) Picture of a fabricated device. (Right) Conceptual drawing of ``hybrid needle design''. The span is the length of the tapered membrane between two silicon needles. (d) ``Infinity design'' is based on the membrane structure. (Left) Picture of a fabricated device. (Right) Conceptual drawing of ``infinity design''. The span is the diameter of the membrane hole.}
\label{fig_6}
\end{figure}

We tested three representative designs based on the alumina membrane ridge waveguides, as follows: (1) straight design, (2) hybrid needle design, and (3) infinity design as shown in Fig.\,\ref{fig_6}. The straight design, as pictured in Fig.\,\ref{fig_6}(b), consists of a fully suspended membrane waveguide across the membrane hole. This provides the least amount of thermal dissipation and provides a relative comparison for the increased performance of the two designs we describe in this paper.  Furthermore, this provides a rough comparison to other works in the literature.  The work of \,\cite{Kimble20} describes a straight waveguide with thermal anchoring structures spaced by 150\,$\mu$m, and is plotted in Fig.\,\ref{fig_6}(a) as a black circle for further reference. It should be noted that this work utilizes silicon nitride ($\rm{n_{\rm{Si_{3}N_{4}}}=2.05}$) as the waveguide material which will have slightly different optical absorption and thermal conductivity compared to alumina ($\rm{n_{\rm{Al_{2}O_{3}}}=1.76}$).  Fig.\,\ref{fig_6} shows the optical power at failure of these three different designs for a range of spans.  The span of each design is defined in Fig.\,\ref{fig_6}(b-d).  Of particular interest, the ``infinity design'' device with the membrane structure was capable of supporting $\sim$ 30\,mW in a vacuum of $5.5 \times 10^{-5}$ torr across a 125\,$\mu$m suspended length in vacuum, respectively (see Fig.\,\ref{fig_5}(d)), which is a record, surpassing other published results by a factor of three\,\cite{Kimble20} and is capable of providing the trap depth of an EFOT more than 1.6\,mK. The ``hybrid needle design'' device with the membrane and needle structures currently manages an optical power of 20.6\,mW in the vacuum over a 250\,$\mu$m suspended length (see Fig.\,\ref{fig_5}(c) and Fig.\,\ref{fig_6}(c)), which is capable of providing the trap depth of an EFOT as $\sim$ 1.2\,mK. These structures have a rather short span and have not yet been proven to support the creation of a MOT surrounding the waveguide. However, transporting several atoms into close proximity of the waveguide is possible with an optical conveyor belt, an optical tweezer, or a surface magnetic trap. Based on these short span devices, we may be able to evanescently trap a single atom with the waveguide by adiabatically raising the potential depth of the EFOT and adiabatically lowering the potential depth of the atom transport trap. In addition, the MOT pictured in Fig.\,\ref{fig_2}(d) is located around the waveguide of this device with span of 400\,$\mu$m. From Fig.\,\ref{fig_6}(a) we estimate that this structure will be capable of handling more than 6mW of optical power. This is capable of providing the trap depth of an EFOT as 350\,$\mu$K, which we believe to be sufficient to trap many atoms directly from cold dense atoms around the integrated photonic platform.

\section{Conclusion}
We demonstrated a record-breaking in-vacuum optical power handling capability of $\sim$30\,mW within the suspended alumina membrane waveguide, which is three times higher than the previously published record with the suspended silicon nitride waveguide\,\cite{Kimble20}. For more than a decade, researchers have explored integrated photonic waveguide platforms to realize trapped neutral atoms via evanescent light and take advantage of the small mode area of the waveguide. This platform enables (1) strong atom-light interaction, (2) enhanced photon collection from trapped atoms through waveguides, (3) low-SWaP operation, and (4) chip-scale design and flexibility. However, integrated photonic waveguides have high absorption loss compared to optical fibers. This makes it extremely difficult to dissipate the heat from optical absorption when operating at the power levels needed to trap the neutral atoms in a vacuum.  Our demonstration of the power handling capability of $\sim$30\,mW should enable sufficient trapping power and lead to a first demonstration of trapped neutral atoms using an integrated, suspended waveguide in vacuum.
	We created a ``hybrid needle design'' with the membrane and needle structures, where the membrane ridge waveguide is suspended over a sub-millimeter center gap in between two undercut silicon needle substrates. This structure efficiently dissipates the heat from the suspended waveguides in vacuum, through the silicon needle substrates, and allows efficient atom loading at close proximity (within 200\,nm) to the waveguides. The transparent alumina membrane film is realized through atomic layer deposition. This photonic atom trap integrated platform (ATIP) can offer chip-level scalability and design flexibility toward a modular and integrated quantum platform. We demonstrated two key milestones: (1) high optical power delivery in vacuum (enough to trap many atoms) and (2) efficient and broadband multi-wavelength guided light. In addition, efficient atom loading into a MOT near nanostructures was demonstrated in previous studies\,\cite{Lee20, Jau17} as well as in this work using our new ``hybrid needle design'' (see Fig.\,\ref{fig_2}(d)). This developed ATIP offers compactness, robustness, manufacturability, and energy efficiency.
	Cold-atom inertial navigation sensors show exceptional sensitivity in the laboratory and have the potential to provide exceptional position, navigation, and timing (PNT) performance in GPS-denied navigation through atom interferometry. The record-breaking in-vacuum optical power handling of the suspended waveguide allows us to develop guided-atom interferometer accelerometers and gyroscopes that can move this exquisite PNT performance out of the lab (decreased SWaP) and into dynamic environments toward real-world missions. This initial demonstration illustrates the potential for further power handling improvements that will also have large implications from atomic sensors to quantum computation, simulation, and networks. This photonic ATIP can also interface atomic spins with integrated photonics and electronics, which will be critical to the function of integrated quantum systems.

\section*{Acknowledgments}
his work was supported by the Laboratory Directed Research and Development program at Sandia National Laboratories and has funding under the DARPA APhI program. Sandia National Laboratories is a multimission laboratory managed and operated by National Technology and Engineering Solutions of Sandia, LLC., a wholly owned subsidiary of Honeywell International, Inc., for the U.S. Department of Energy's National Nuclear Security Administration under contract DE-NA-0003525. This paper describes objective technical results and analysis. Any subjective views or opinions that might be expressed in the paper do not necessarily represent the views of the U.S. Department of Energy or the United States Government.' 

\section*{Disclosures}
The authors declare no conflicts of interest.

\section*{Data availability}
Data underlying the results presented in this paper are not publicly available at this time but may be obtained from the authors upon reasonable request.

\medskip

$^{\dagger}$ M. Gehl and W. Kindel contributed equally to this work.
 
 \medskip


\end{document}